\documentclass[twocolumn, prb, superscriptaddress]{revtex4-1}
\usepackage{graphicx}
\usepackage{amsmath}
\usepackage{hyperref}

\newcommand{\mof}{MOF-74}
\newcommand{\reaction}{H$_2$O$\;\rightarrow\;$OH+H}
\newcommand{\reactionD}{D$_2$O$\;\rightarrow\;$OD+D}

\newcommand{\reactionFA}{H$_2$O+CO$\;\rightarrow\;$HCO$_2$H}

\newcommand{\reactionHFA}{OH+H+CO$\;\rightarrow\;$HCO$_2$H}

\graphicspath{{./figures/}}

\newcommand{\UTD}{Department of Materials Science and Engineering,
University of Texas at Dallas, Richardson, Texas 75080, USA}
\newcommand{\WFU}{Department of Physics, Wake Forest University,
Winston-Salem, North Carolina 27109, USA} 
\newcommand{\RU}{Department of
Chemistry and Chemical Biology, Rutgers University, Piscataway, New
Jersey 08854, USA}
\begin{document}
%%%%%%%%%%%%%%%%%%%%%%%%%%%%%%%%%%%%%%%%%%%%%%%%%%%%%%%%%%%%%%%%%%%%%%%%

%%%%%%%%%%%%%%%%%%%%%%%%%%%%%%%%%%%%%%%%%%%%%%%%%%%%%%%%%%%%%%%%%%%%%%%%
\title{Chemistry in confined spaces:
Reactivity of the Zn-MOF-74 channels}

\author{S. Zuluaga}                               \affiliation{\WFU}
\author{E. M. A. Fuentes-Fernandez}               \affiliation{\UTD}
\author{K. Tan}                                   \affiliation{\UTD}
\author{C. A. Arter}                              \affiliation{\WFU}
\author{J. Li}                                    \affiliation{\RU}
\author{I. J. Chabal}                             \affiliation{\UTD}
\author{T. Thonhauser} \email{thonhauser@wfu.edu} \affiliation{\WFU}

\date{\today}

\begin{abstract}
Using infrared spectroscopy combined with \emph{ab initio} methods we
study reactions of H$_2$O and CO inside the confined spaces of Zn-MOF-74
channels. Our results show that, once the water dissociation reaction
{\reaction} takes place at the metal centers, the addition of 40~Torr of
CO at 200~$^{\circ}$C starts the production of formic acid via
{\reactionHFA}. Our detailed analysis shows that the overall reaction
{\reactionFA} takes place in the confinement of {\mof} without an
external catalyst, unlike the same reaction on flat surfaces. This
discovery has several important consequences: It opens the door to a new
set of catalytic reactions inside the channels of the {\mof} system, it
suggests that a recovery of the MOF's adsorption capacity is possible
after it has been exposed to water (which in turn stabilizes its crystal
structure), and it produces the important industrial feedstock formic
acid.
\end{abstract}

\maketitle
%%%%%%%%%%%%%%%%%%%%%%%%%%%%%%%%%%%%%%%%%%%%%%%%%%%%%%%%%%%%%%%%%%%%%%%%


%%%%%%%%%%%%%%%%%%%%%%%%%%%%%%%%%%%%%%%%%%%%%%%%%%%%%%%%%%%%%%%%%%%%%
\section{Introduction}
%%%%%%%%%%%%%%%%%%%%%%%%%%%%%%%%%%%%%%%%%%%%%%%%%%%%%%%%%%%%%%%%%%%%%

Metal organic framework (MOF) materials are porous crystals widely
studied for important applications and industrial processes such as gas
storage and sequestration,\cite{Liu_2012:progress_adsorption-based,
Murray_2009:hydrogen_storage, Li_2011:carbon_dioxide,
Qiu_2009:molecular_engineering, Nijem_2012:tuning_gate,
Lee_2015:small-molecule_adsorption, Zhao_2008:current_status,
Rosi_2003:hydrogen_storage, Wu_2012:commensurate_adsorption,
He_2014:methane_storage} molecular
sensing,\cite{Kreno_2012:metal-organic_framework,
Canepa_2015:structural_elastic, Serre_2007:role_solvent-host,
Allendorf_2008:stress-induced_chemical, Tan_2011:mechanical_properties,
Hu_2014:luminescent_metal-organic}
polymerization,\cite{Uemura_2009:polymerization_reactions,
Vitorino_2009:lanthanide_metal}
luminescence,\cite{Allendorf_2009:luminescent_metal,
White_2009:near-infrared_luminescent} non-linear
optics,\cite{Bordiga_2004:electronic_vibrational} magnetic
networks,\cite{Kurmoo_2009:magnetic_metal-organic} targeted drug
delivery,\cite{Horcajada_2010:porous_metal-organic-framework}
multiferroics,\cite{Stroppa_2011:electric_control,
Stroppa_2013:hybrid_improper, Di-Sante_2013:tuning_ferroelectric} and
catalysis.\cite{Wu_2007:heterogeneous_asymmetric,
Lee_2009:metal-organic_framework, Zou_2006:preparation_adsorption,
Luz_2010:bridging_homogeneous} In particular, {\mof}
[$\mathcal{M}_2$(dobdc), $\mathcal{M}$ = Mg$^{2+}$, Zn$^{2+}$,
Ni$^{2+}$, Co$^{2+}$, and dobdc=2,5-dihydroxybenzenedicarboxylic acid]
has shown great potential for the adsorption of small molecules such as
H$_2$,\cite{Liu_2008:increasing_density, Zhou_2008:enhanced_h2} CO$_2$,
\cite{Wu_2010:adsorption_sites, Dietzel_2008:adsorption_properties,
Caskey_2008:dramatic_tuning} N$_2$,
\cite{Valenzano_2010:computational_experimental} and CH$_4$,
\cite{Wu_2009:high-capacity_methane} among others. 

\begin{figure}[t]
\centering\reflectbox{\includegraphics[width=0.9\columnwidth]{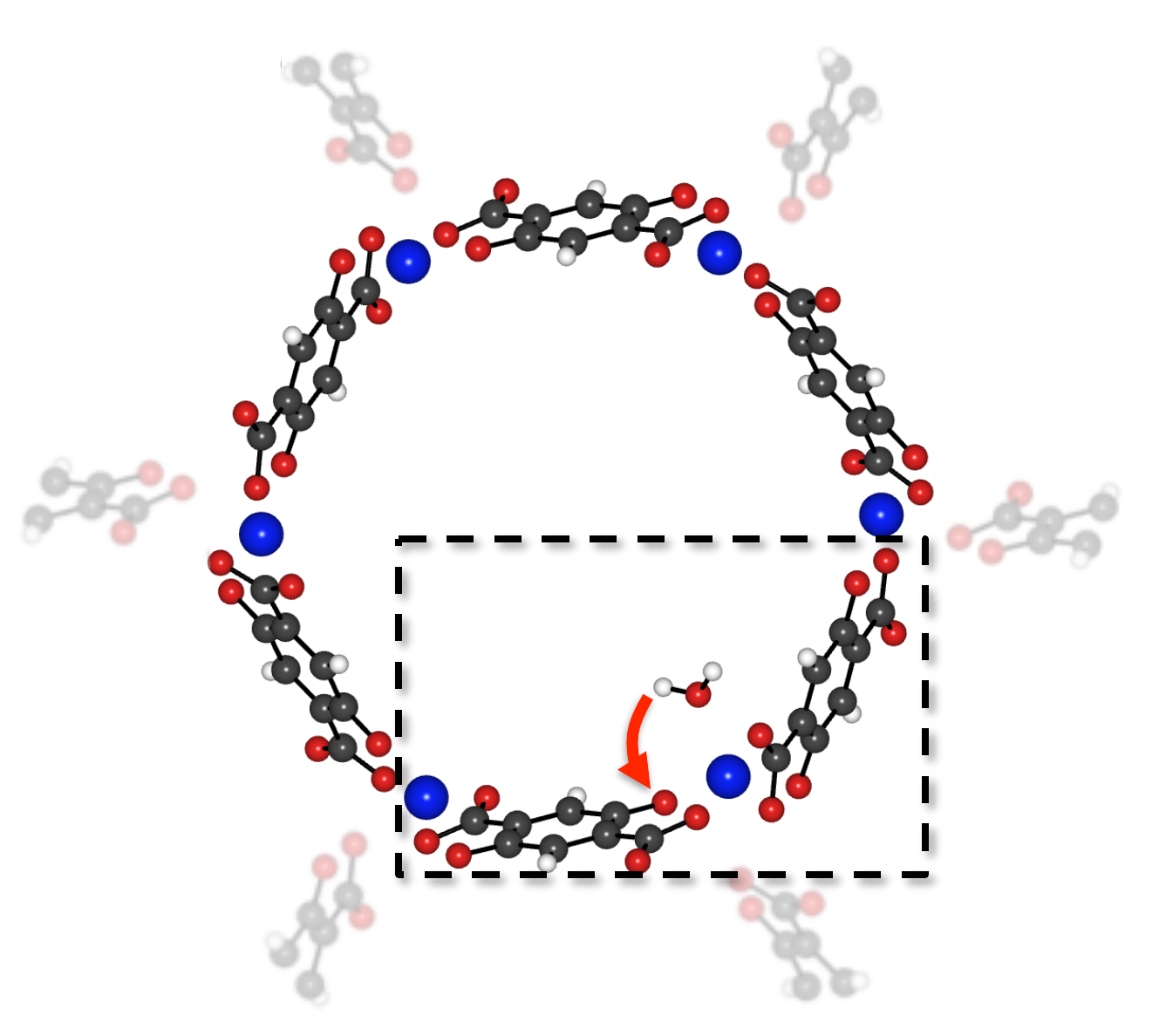}}
\caption{\label{fig:MOF-74} Zn-MOF-74 with its hexagonal channels
clearly visible. The open-metal sites at the corners form the primary
adsorption sites. The arrow indicates how the H of the water is
transferred to the O of the linker during the H$_2$O~$\rightarrow$~OH+H
reaction. Black, red, white, and blue spheres represent C, O, H, and Zn
atoms. The box shows the portion of MOF-74 visible in other figures,
albeit from a slightly different angle.}
\end{figure}

The favorable reactivity of {\mof} has been widely
studied.\cite{Valvekens_2014:metal-dioxidoterephthalate_mofs,
Yao_2014:cpo-27-m_heterogeneous, Kim_2015:low-temperature_co,
Sun_2015:mixed-metal_strategy, Zhang_2015:synthesis_nanosized,
Ruano_2015:nanocrystalline_m-mof-74} For example, Co-{\mof} exhibits a
catalytic activity towards CO
oxidation,\cite{Kim_2015:low-temperature_co} originating from the high
density of lewis acidic coordinatively unsaturated sites and the MOF's
porosity. The inclusion of Co atoms into Ni-{\mof} results in a mixed
system (Co/Ni-{\mof}) that shows activity towards the oxidation of
cyclohexene, where the catalytic performance of the mixed system is
higher than the one of pure
Co-{\mof}.\cite{Sun_2015:mixed-metal_strategy} On the other hand, our
previous results have shown that several members of the {\mof} family are
able to catalyze the dissociation of water into H and OH groups
({\reaction}, see Fig.~\ref{fig:MOF-74}) at low temperatures and
pressures, i.e.\ above 150~$^{\circ}$C and at 8~Torr of
H$_2$O.\cite{Zuluaga_2016:understanding_controlling,
Tan_2014:water_reaction, Tan_2015:water_interactions}  This particular
catalytic reaction is responsible for the loss of crystal structure and
adsorption capacity after exposure of {\mof} to
water,\cite{Zuluaga_2016:understanding_controlling} and constitutes one
of the main hurdles for wide-spread applications of MOFs in general and
MOF-74 in particular.  This challenge has motivated our efforts to look
for new catalytic reactions inside the confined channels of {\mof},
further reacting the undesirable products of the {\reaction} reaction in
order to overcome these hurdles.

In this work, we show that introducing CO molecules into the pores of
{\mof}---after the {\reaction} reaction has taken place---enables the
reaction \reactionHFA. Our results show that the overall reaction
{\reactionFA} takes place in the confinement of the MOF without an
external catalyst, with a number of important consequences: First, it
showcases the reactivity inside the well-controlled and isolated
environment of the \mof\ channels.  This aspect is very important, as the
confinement of the \mof\ environment catalyzes reactions that would
otherwise require very high pressure, bringing significant
simplifications for experiments and possible MOF applications.  Next, it
shows initial indications of a partial adsorption capacity recovery after
exposure of MOF to water, as the OH groups that otherwise poison the
metal centers are now bound to and removed as formic acid.  In turn, it
increases the crystal structure stability of \mof\ by removing the OH and
H groups that cause the instability (note that, due to their strong
binding, those groups cannot be removed by thermal
activation).\cite{Zuluaga_2016:understanding_controlling} And finally, it
binds the toxic CO and produces formic acid, a non-toxic liquid with
4.4~wt\% hydrogen and thus a promising hydrogen carrier
\cite{Gu_2011:synergistic_catalysis, Tedsree_2011:hydrogen_production,
Zell_2013:efficient_hydrogen, Zhang_2013:monodisperse_agpd,
Joo_2008:breakthroughs_hydrogen} and an important feedstock
medical/industrial chemical.  The use of Pd as catalytic material in
direct formic acid fuel cells has brought interesting developments in
this area,\cite{Ha_2006:characterization_high,
Chang_2014:effective_pd-ni2pc, Shen_2013:improvement_mechanism,
Feng_2011:high_activity} highlighting formic acid as a valuable asset for
a hydrogen economy.

%%%%%%%%%%%%%%%%%%%%%%%%%%%%%%%%%%%%%%%%%%%%%%%%%%%%%%%%%%%%%%%%%%%%%
\section{Experimental and Theoretical Methods}
%%%%%%%%%%%%%%%%%%%%%%%%%%%%%%%%%%%%%%%%%%%%%%%%%%%%%%%%%%%%%%%%%%%%%

%%%%%%%%%%%%%%%%%%%%%%%%%%%%%%%%%%%%%%%%%%%%%%%%%%%%%%%%%%%%%%%%%%%%%
\subsection{Zn-{\mof}}

Out of the isostructural $\mathcal{M}$-\mof\ family, Zn-{\mof} exhibits the
highest catalytic activity towards the {\reaction}
reaction.\cite{Tan_2014:water_reaction} We thus use this system to study the
{\reactionFA} reaction through a combination of \emph{ab initio} simulations
and experiments.

%%%%%%%%%%%%%%%%%%%%%%%%%%%%%%%%%%%%%%%%%%%%%%%%%%%%%%%%%%%%%%%%%%%%%
\subsection{Hydrogen vs. Deuterium}\label{sec:hydrogen_vs_deuterium}

Only recently, our work showed direct evidence of the water dissociation
reaction {\reaction} at the metal centers of {\mof} above
150~$^{\circ}$C.\cite{Tan_2014:water_reaction,
Tan_2015:water_interactions} In this reaction, the water first binds to
an open-metal site and then donates one H to the nearby O at the linker;
the remaining OH group stays at the open-metal site, see
Fig.~\ref{fig:MOF-74}.  Interestingly, this reaction can only be observed
when heavy water D$_2$O is used. Its fingerprint is a sharp peak at
970~cm$^{-1}$ in the IR spectrum, corresponding to the O--D vibration at
the linker.\cite{Tan_2014:water_reaction} When H$_2$O is used instead,
the peak appears at a higher frequency, where it couples with and is
masked by the vibrational modes of the MOF and becomes impossible to
detect.  Therefore, the main focus of our experiments is on the water
reaction with D$_2$O. We refer to the resulting deuterated formic acid as
FA(D).  Nonetheless, we do show that the reaction also occurs with
H$_2$O, referring to the resulting formic acid as FA(H). For simplicity,
throughout the text we may generally say \emph{water}, even when
experiments are done with \emph{heavy water}.

%%%%%%%%%%%%%%%%%%%%%%%%%%%%%%%%%%%%%%%%%%%%%%%%%%%%%%%%%%%%%%%%%%%%%
\subsection{Experimental Details and Procedure}\label{sec:exp_procedure}

Our experiments are divided into 3 steps:

\textbf{(i) Preparation and activation of the sample:} Zn-{\mof} powder
($\sim$2~mg) was pressed onto a KBr pellet ($\sim$1~cm diameter, 1--2~mm
thick). The sample was placed into a high-pressure high-temperature cell
(product number P/N 5850c, Specac Ltd, UK) at the focal point of an infrared
spectrometer (Nicolet 6700, Thermo Scientific, US). The sample was activated
under vacuum at 180~$^{\circ}$C for 4 hours and then cooled down to room
temperature to measure CO$_2$ absorption by introducing 6~Torr of CO$_2$ into
the cell until saturation (30 minutes). Then, the area under the peak at
2338~cm$^{-1}$ was determined, which is a characteristic peak of CO$_2$
adsorbed on the Zn site and thus a quantitative measure of the CO$_2$
uptake.\cite{Yao_2012:analyzing_frequency} Thereafter, the cell was evacuated
under vacuum ($<$~20~mTorr) at a temperature of 150~$^{\circ}$C for a period of
4 hours.

\textbf{(ii) Dissociation reaction:} The sample was heated to
200~$^{\circ}$C. 8 Torr of D$_2$O were then introduced into the cell
until saturation occurred (8~hours) to start the dissociation reaction.
Spectra were recorded as a function of time during the adsorption
process to evaluate the 970 cm$^{-1}$ peak, i.e.\ the fingerprint of the
{\reactionD} reaction. Thereafter, evacuation under vacuum
($<$~20~mTorr) for a period of 4~hours at 150~$^{\circ}$C was required
to evacuate  the water gas phase completely and avoid further reaction.
Note that this temperature is not high enough to also remove the OD and
D products of the dissociation reaction.  Then, at room temperature,
CO$_2$ adsorption was measured again and the cell was evacuated as in
step (i).

\textbf{(iii) Formic acid production and removal:} The temperature in
the cell was raised back to 200~$^{\circ}$C and 40~Torr of CO were
introduced for 1 hour to start the formic acid production. Spectra were
recorded. Thereafter, the cell was evacuated for 3 hours under vacuum
($<$~20~mTorr) at 200~$^{\circ}$C, removing the formic acid and
unreacted CO, while spectra were recorded. Then, CO$_2$ adsorption at
room temperature was measured and the cell was evacuated as in step (i).
This production and removal step was repeated two times and we refer to
each occurrence as \emph{removal~1} and \emph{removal~2}.

%%%%%%%%%%%%%%%%%%%%%%%%%%%%%%%%%%%%%%%%%%%%%%%%%%%%%%%%%%%%%%%%%%%%%
\subsection{Computational Details}

\emph{Ab initio} modeling was performed at the density functional theory level,
using \textsc{quantum espresso}\cite{Giannozzi_2009:quantum_espresso} with the
vdW-DF functional.\cite{Thonhauser_2015:spin_signature, Berland_2015:van_waals,
Langreth_2009:density_functional, Thonhauser_2007:van_waals} Ultrasoft pseudo
potentials were used with cutoffs of 544~eV and 5440~eV for the wave functions
and charge density. Due to the large dimensions of the unit cell, only the
$\Gamma$-point was used. During relaxations all atom positions were optimized
until forces were less than 2.6$\times$10$^{-4}$~eV/\AA. Reaction barriers were
found with a transition-state search algorithm, i.e.\ the climbing-image
nudged-elastic band method.\cite{Henkelman_2000:climbing_image,
Henkelman_2000:improved_tangent} The primitive cell of our pristine Zn-{\mof}
system contained 54 atoms and has space group R$\bar{3}$.  Additional
atoms/molecules were added as appropriate for the reactants.  The rhombohedral
axes are $a=b=c=15.105$~\AA\ and
$\alpha=\beta=\gamma=117.78^\circ$.\cite{Zhou_2008:enhanced_h2}

%%%%%%%%%%%%%%%%%%%%%%%%%%%%%%%%%%%%%%%%%%%%%%%%%%%%%%%%%%%%%%%%%%%%%
\section{Results and Discussion}
%%%%%%%%%%%%%%%%%%%%%%%%%%%%%%%%%%%%%%%%%%%%%%%%%%%%%%%%%%%%%%%%%%%%%

%%%%%%%%%%%%%%%%%%%%%%%%%%%%%%%%%%%%%%%%%%%%%%%%%%%%%%%%%%%%%%%%%%%%%
\subsection{Confirming Formic Acid Production and Removal}

\begin{figure}[t]
\includegraphics[width=\columnwidth]{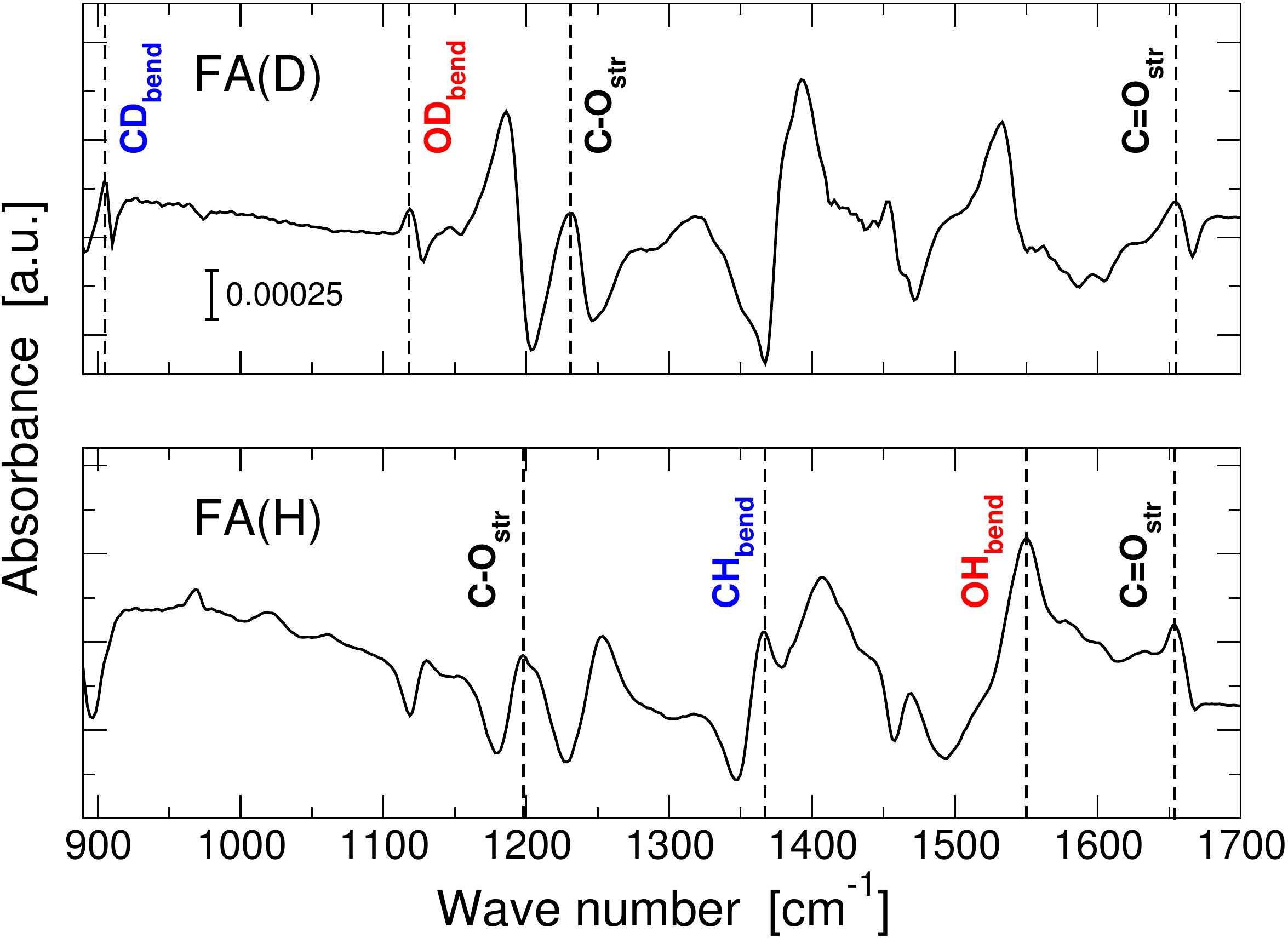}
\caption{\label{formic_acid_fig} IR absorption spectra of Zn-{\mof}
reacted with D$_2$O (top panel) and H$_2$O (bottom panel) followed by
the addition of 40~Torr of CO for 1 hour at 200~$^{\circ}$C. The panels
show the characteristic peaks of FA(D) and FA(H). Both samples are
referenced to pure Zn-{\mof}.}
\end{figure}

We begin by showing experimental evidence that the reactive environment inside
the \mof\ channels catalyses the formic acid production (through water
dissociation) \reactionHFA\ after the water dissociation \reaction\ has taken
place.  To this end, we follow the three-step procedure outlined in
Sec.~\ref{sec:exp_procedure}. After the introduction of CO in step (iii), our
IR spectra in Fig.~\ref{formic_acid_fig} clearly show the presence of FA(D) and
FA(H) molecules.\cite{Millikan_1958:infrared_spectra}  As expected, due to the
deuterium presence, the FA(D) peaks (CD$_{\text{bend}}$ and OD$_{\text{bend}}$)
are red shifted with respect to FA(H) peaks (CH$_{\text{bend}}$ and
OH$_{\text{bend}}$) by a factor of $\sim$1.4.  C--O and C=O modes are less
disturbed (shifted), as they are not directly affected by the presence of
deuterium or hydrogen. The OH$_{\text{bend}}$ vibrational mode signal at
$\sim$1550~cm$^{-1}$ for FA(H) appears very close to a strong MOF mode at
$\sim$1530~cm$^{-1}$, and this vibrational mode may be contributing to the
OH$_{\text{bend}}$ signal.  On the other hand, the signal at $\sim$1530
cm$^{-1}$ in the FA(D) spectrum may be due to a hydrogen contamination of the
deuterated water, increased by the vibrations of the MOF modes.  

\begin{figure}[t]
\includegraphics[width=\columnwidth]{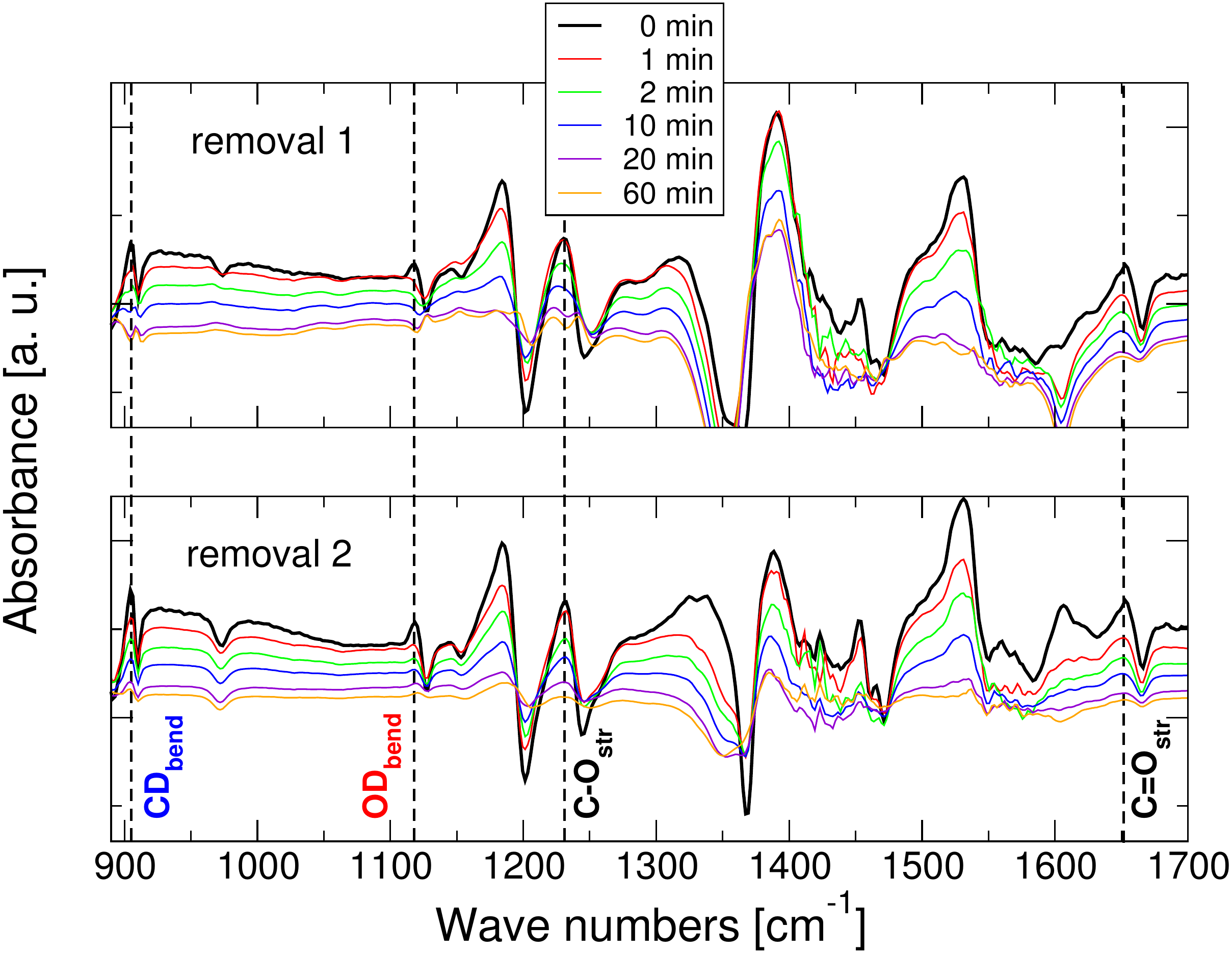}
\caption{\label{desoprtion_fig} IR spectra of the desorption of FA(D) as
a function of time for \emph{removal 1} and \emph{removal 2}. Both
figures are referenced to pure Zn-{\mof}.}
\end{figure}

In Fig.~\ref{desoprtion_fig} we show how the characteristic peaks of
FA(D) disappear as a function of time during the removal in step (iii),
showing that the produced formic acid can readily be removed. Note that
these experiments rely on the detection of the linker O--D mode at
970~cm$^{-1}$ and are thus only performed for the deuterated case (see
Sec.~\ref{sec:hydrogen_vs_deuterium}). We will henceforth only discuss
the deuterated case. It is interesting to note
that---while the starting point for \emph{removal 1} and \emph{2} are
comparable to within 6\%---the desorption becomes faster.  For example,
in the former case 35\% of FA(D) was removed after 20 min, while in the
latter 62\% was removed during the same time. This fact, together with
the fact that several removals are necessary to react all OD and D
groups suggests a bottleneck in diffusion of the reactants and products,
discussed further below.

\begin{figure}[t]
\includegraphics[width=1.0\columnwidth]{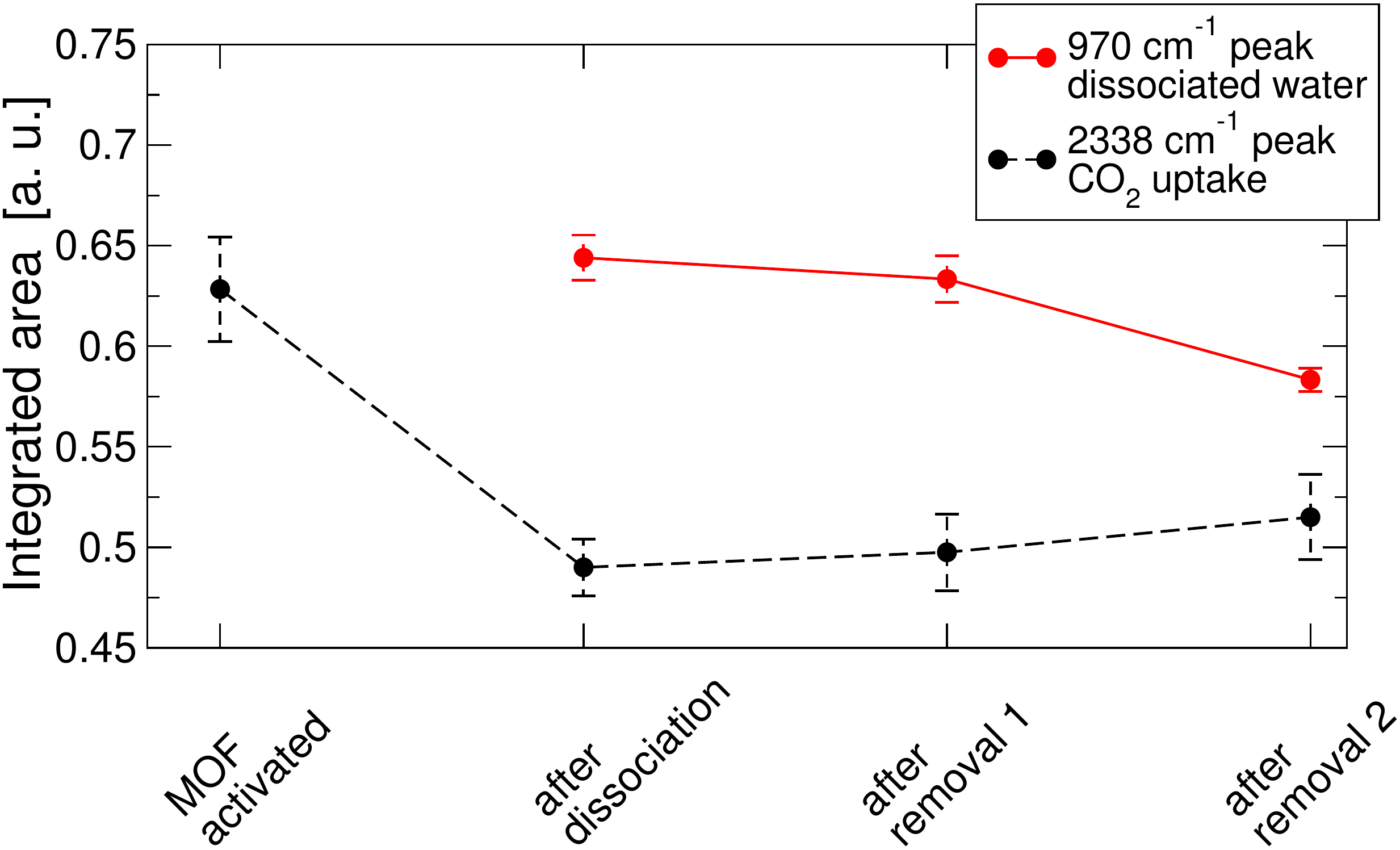}
\caption{\label{CO2_970_fig} Integrated areas of the 970~cm$^{-1}$ peak (a measure
of the amount of dissociated water) and 2338~cm$^{-1}$ peak
(a measure of the CO$_2$ uptake capacity).}
\end{figure}

After the water dissociation reaction happens, its products (OD or OH)
are strongly bound to primary adsorption site in the MOF and take up
valuable adsorption sites. This undesirable decrease of the MOF's
adsorption capacity is well known\cite{Tan_2014:water_reaction,
Tan_2015:water_interactions, Zuluaga_2016:controlling_water} and
unfortunately limits the applicability of MOF materials to non-humid
environments. Note that the water dissociation products bind so strongly
to the MOF that a simple removal through activation is not possible
before the MOF disintegrates.  Other means to recover the uptake capacity
of MOFs after exposure to water are thus highly desirable.  Our
production and removal of formic acid reacts those unwanted groups that
are otherwise bound to the MOF after the water dissociation reaction.  We
now show that this process also partially restores the MOF's
small-molecule uptake capacity. In Fig.~\ref{CO2_970_fig} we track the
970~cm$^{-1}$ peak (a measure for the amount of dissociated heavy water
present in the MOF cavity)\cite{Tan_2014:water_reaction} as well as the
2338~cm$^{-1}$ peak (a measurement of the CO$_2$ adsorption
capacity)\cite{Yao_2012:analyzing_frequency} at different stages of our
experiment. We see that the former decreases as we introduce CO into the
system, i.e\ by 1.6\% after \emph{removal 1} and 7.9\% after
\emph{removal 2}. This confirms that we have successfully removed the D
groups from the linkers of the MOF. On the other hand, the latter---after
an expected big reduction in the CO$_2$ uptake capacity after the D$_2$O
dissociation (22\%)---increases by 1.5\% and 5.1\% after \emph{removal 1}
and \emph{2}. While the MOF's uptake capacity recovery is relatively
small per removal cycle, our results constitute the first
proof-of-principle that such a recovery is even possible.

As expected, the decrease in the amount of dissociated water (area under
the peak at 970~cm$^{-1}$) goes hand-in-hand with the increase of the
CO$_2$ uptake capacity (area under the peak at 2338~cm$^{-1}$). However,
it is interesting to see that more than one removal cycle is necessary to
restore a significant amount of uptake capacity. In principle, the
partial pressure of 40 Torr CO introduced into the system should be more
than enough (we estimate that it results in at least 6 CO molecules per
unit cell) to react all OD and D groups. However, this is not the case,
see Fig~\ref{CO2_970_fig}. We conclude that the produced formic acid
inhibits diffusion of CO deeper into the bulk. After each removal of
formic acid and the renewed introduction of CO, the process picks up
where it had left off earlier, working from the MOF surface into the bulk
until, eventually, all OD and D groups have been reacted.  Work to reduce
the number of cycles thus needs to focus on diffusion in \mof\
\cite{Canepa_2013:diffusion_small, Lee_2012:preparation_ni-mof-74,
Lin_2013:understanding_co2} as well as using similar reactions with
different products.

\begin{figure}[t]
\includegraphics[width=1.0\columnwidth]{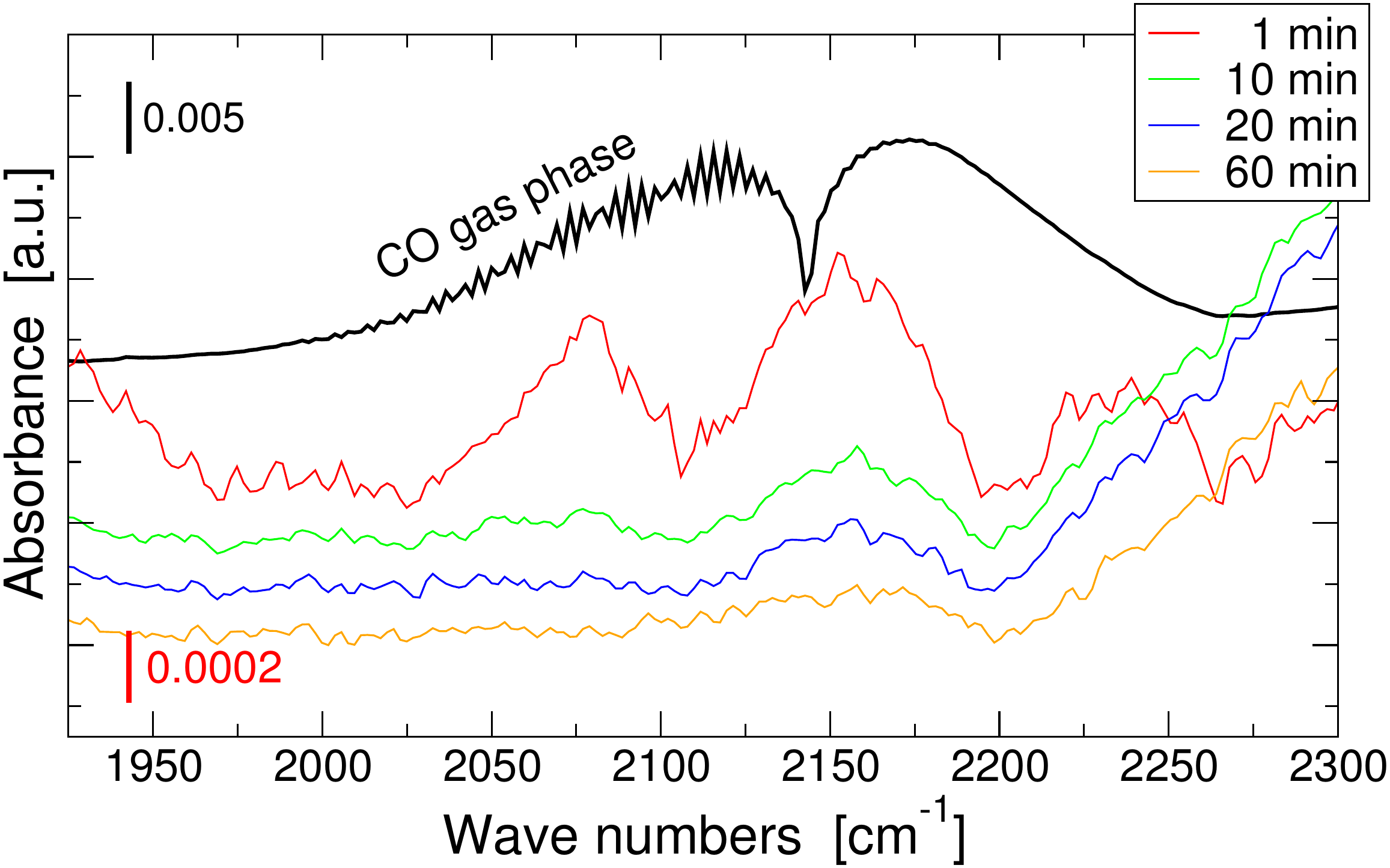}
\caption{\label{CO_fig} CO region of the IR spectrum during \emph{removal~1}. 
The black line is taken just before CO evacuation and is measured on a scale
of 0.005, since the CO gas-phase signal is very strong.
Thereafter, IR spectra are taken at 1, 10, 20, and 60 minutes during evacuation,
measured on the smaller scale of 0.0002.}
\end{figure}

The CO region in the IR spectrum also provides information on the
mechanism of formic acid formation. Figure~\ref{CO_fig} shows the CO
region during \emph{removal~1} at several stages. When CO gas is still
inside the MOF, the predominant signal is at 2173~cm$^{-1}$.
However, after 1 min of
desorption, a shift to lower frequencies (2150~cm$^{-1}$) is observed. 
This indicates that the majority of the CO gas phase has been evacuated
and now the IR spectrum is dominated by a signal that suggests a stronger
interaction between the CO and the MOF, such as in the CO$_2$H+H state
(see Fig.~\ref{fig:states}).  After longer periods of desorption, the
intensity of the signal is reduced as the chamber is evacuated.

Before we continue to study the nature of the reaction, we give an
estimate of how much formic acid is produced. From our calculations we
know that the crystal density of
Zn-MOF-74 is 1.231~g/mL, with a volume of 3944.65~\AA$^3$ for the
hexagonal cell (note that the hexagonal cell contains 18 metal centers
and is three times bigger than the rhombohedral
representation).\cite{Zhou_2008:enhanced_h2} Based on that, we calculate
that in our sample (2~mg of Zn-MOF-74) we have 4.12$\times$10$^{17}$
hexagonal unit cells. According to Fig.~\ref{CO2_970_fig}, we observe a
reduction of 22\% in the CO$_2$ adsorption capacity, suggesting that we
produced $\sim$4~OD+D groups every 18 metal centers. Therefore, when CO is
introduced into the cell and 5.1\% of the CO$_2$ adsorption capacity is
recovered, we estimate a production of 3.95$\times$10$^{17}$ formic acid
molecules. This corresponds to 2.311$\times$10$^{-5}$~mL of formic acid
in the 2~mg of Zn-{\mof}, or 11.55~$\mu$L/g$_\text{MOF}$. Clearly, this is a
small quantity, but as mention before, our goal is to investigate the
chemistry in the confined spaces of Zn-MOF-74 and not the mass production
of formic acid.

%%%%%%%%%%%%%%%%%%%%%%%%%%%%%%%%%%%%%%%%%%%%%%%%%%%%%%%%%%%%%%%%%%%%%
\subsection{Pathway of the Formic Acid Reaction}

\begin{figure*}
\reflectbox{\includegraphics[width=0.49\columnwidth]{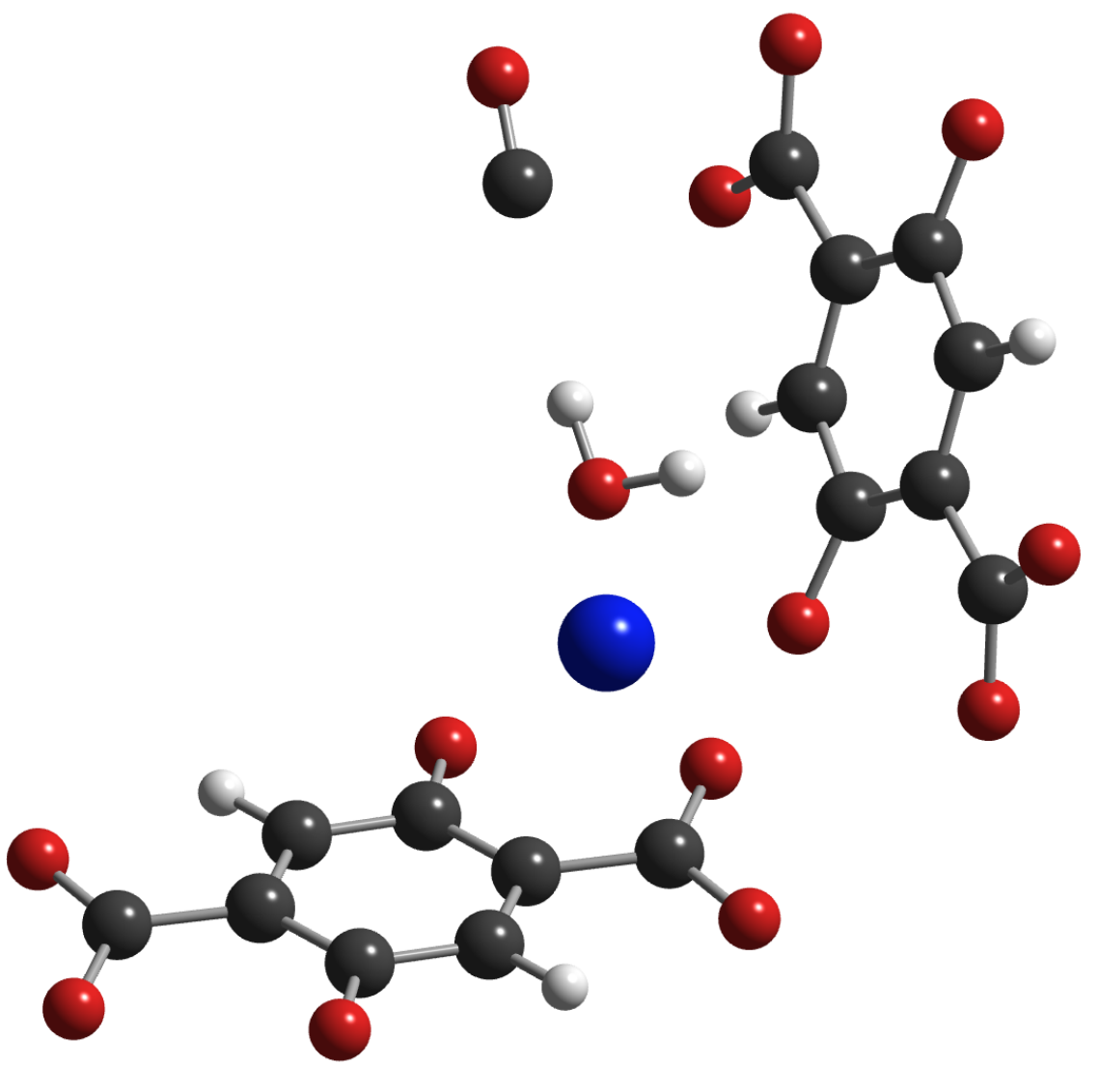}\hfill}
\reflectbox{\includegraphics[width=0.49\columnwidth]{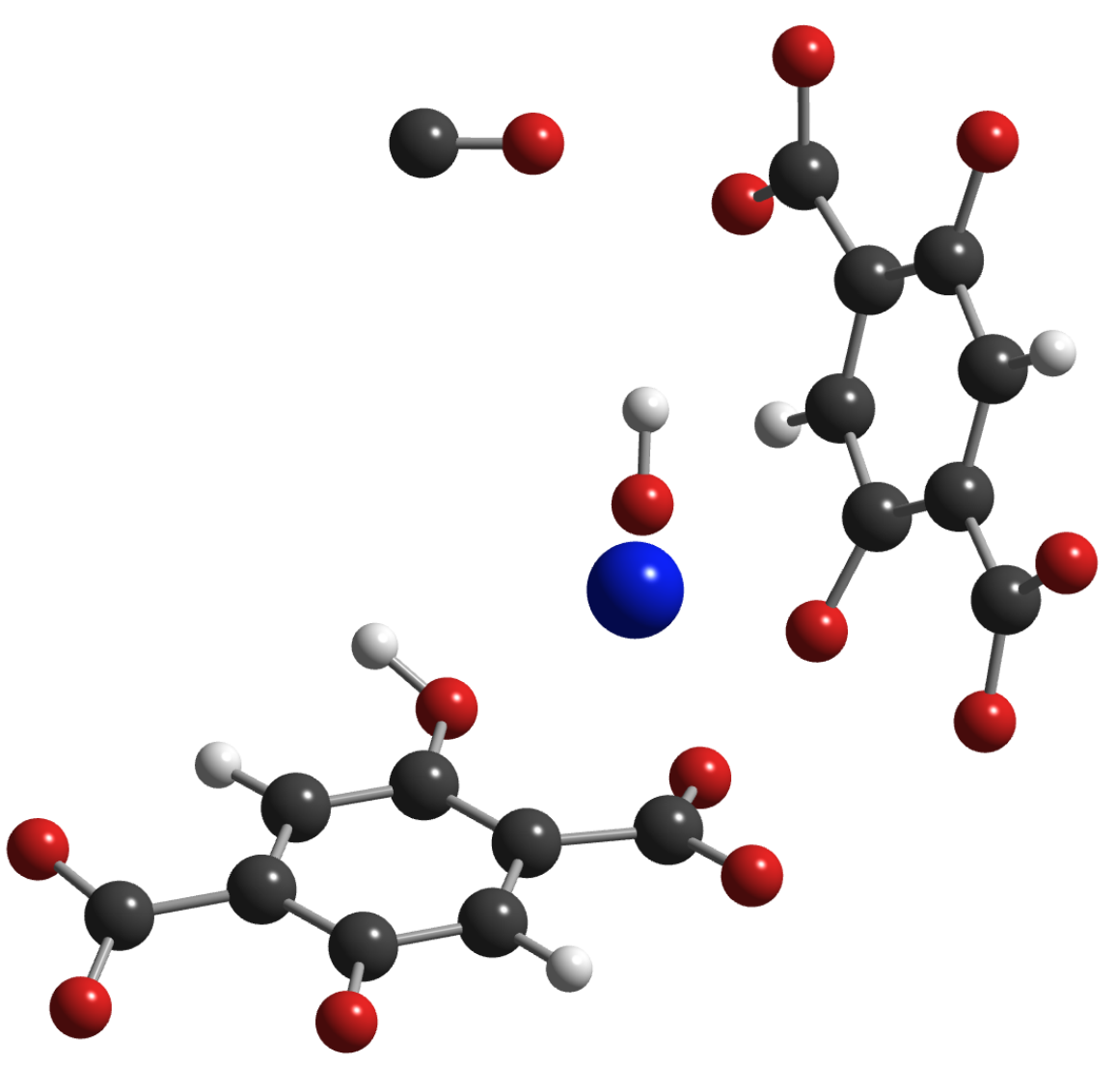}\hfill}
\reflectbox{\includegraphics[width=0.49\columnwidth]{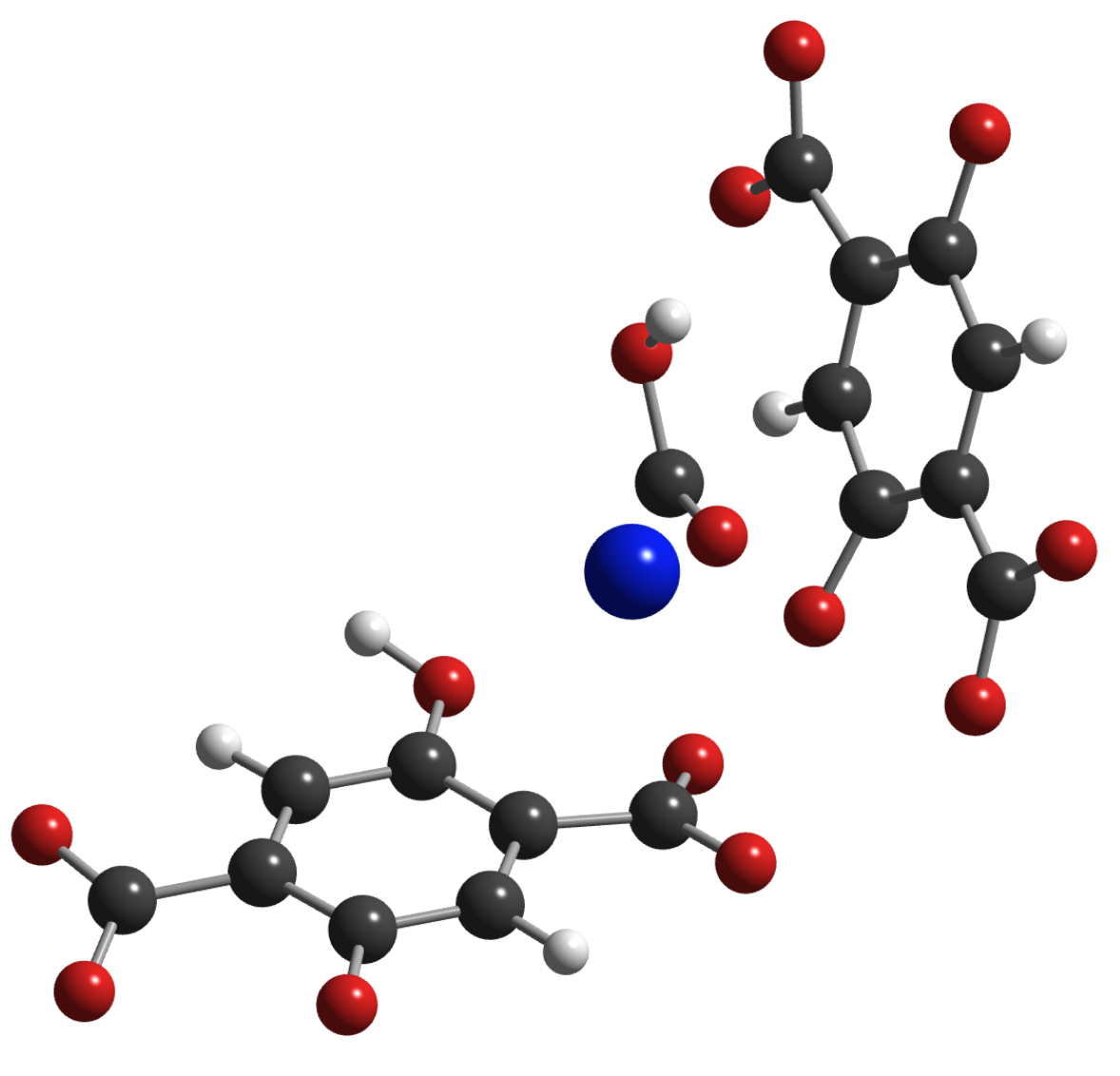}\hfill}
\reflectbox{\includegraphics[width=0.49\columnwidth]{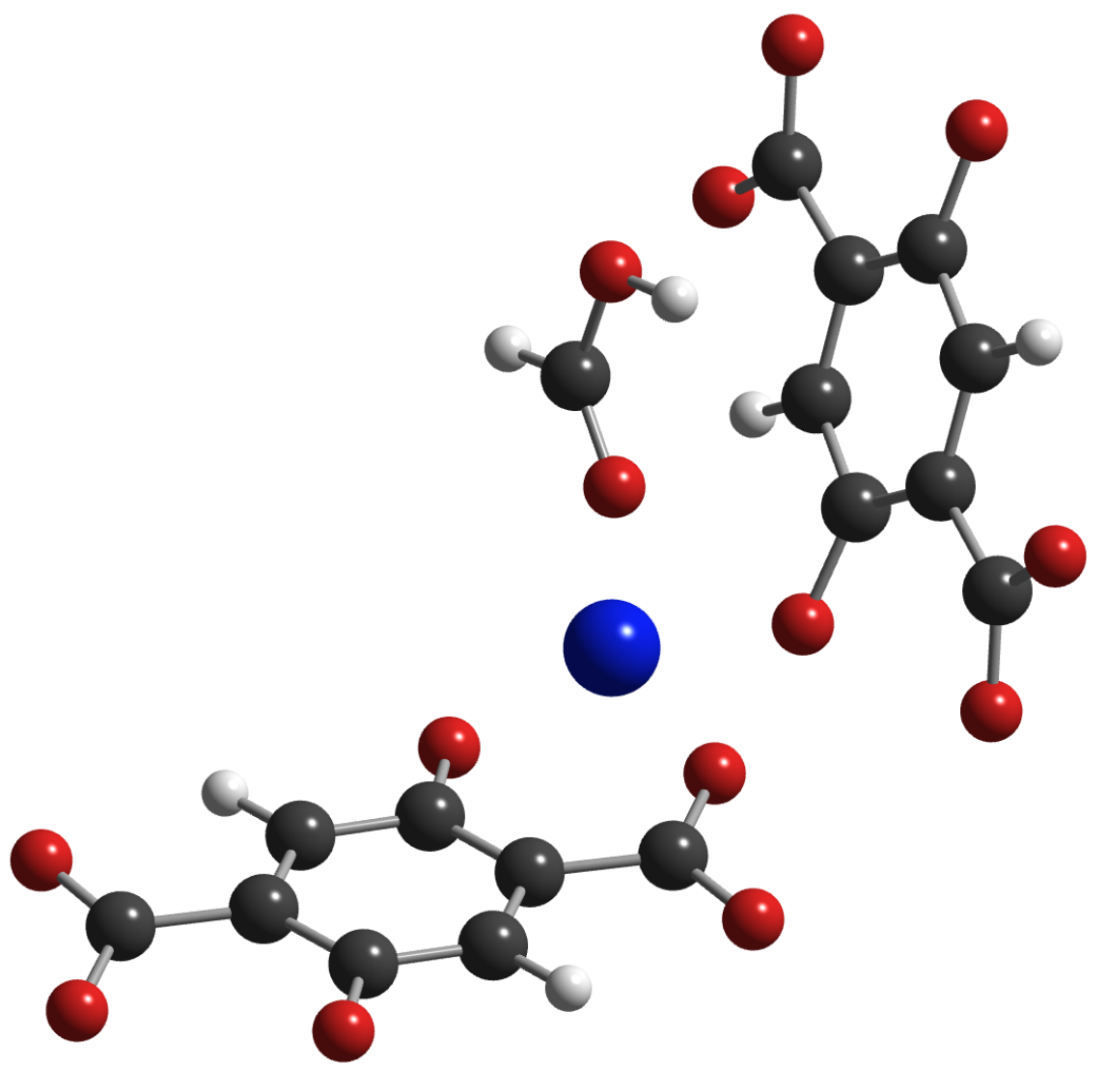}\hfill}
\hspace*{\fill}H$_2$O+CO\hfill$\longrightarrow$\hfill OH+H+CO
\hfill$\longrightarrow$\hfill CO$_2$H+H \hfill$\longrightarrow$\hfill
HCO$_2$H\hspace*{\fill}\mbox{}
\caption{\label{fig:states} Structures for the reactants (H$_2$O+CO),
stable states (OH+H+CO, CO$_2$H+H), and products (HCO$_2$H). First, the
water is adsorbed at the open-metal site, while CO is adsorbed at a
secondary site.  Then, the water dissociates into OH+H; OH remains at
the open-metal site and H is transferred to the O at the linker; CO is
still adsorbed at the secondary site. Then, CO reacts with OH to form
CO$_2$H at the metal center. Finally, the H from the linker reacts with
the CO$_2$H to form HCO$_2$H at the open-metal site.}
\end{figure*}

We now investigate the nature of the formic acid reaction {\reactionFA}
and give and explanation of how it takes place. We know that the first
step is the dissociation of water at the metal centers {\reaction},
which we have studied in detail before.\cite{Tan_2014:water_reaction,
Zuluaga_2016:understanding_controlling, Zuluaga_2016:controlling_water}
We find that the water dissociation takes place above 150~$^\circ$C with
an energy barrier up to 1.09~eV, depending on the number of water
molecules involved in the reaction.\cite{Zuluaga_2016:controlling_water}
The second step of the reaction starts by the introduction of CO at
200~$^\circ$C, which catalyses the {\reactionHFA} reaction.  Based on
this information, and taking into account that the metal centers are
poisoned by the OH groups after the {\reaction} reaction, we propose the
following mechanism for the overall reaction: Once the {\reaction}
reaction takes place, the added CO molecules interact with the OH groups
at the metal centers to form CO$_2$H adsorbed at the metal center.
Thereafter, the CO$_2$H molecule interacts with the H at the linker to
form formic acid HCO$_2$H. Overall, the reaction pathway follows
H$_2$O+CO $\rightarrow$ OH+H+CO $\rightarrow$ CO$_2$H+H $\rightarrow$
HCO$_2$H, as depicted in Fig.~\ref{fig:states}.

\begin{figure}[t]
\includegraphics[width=\columnwidth]{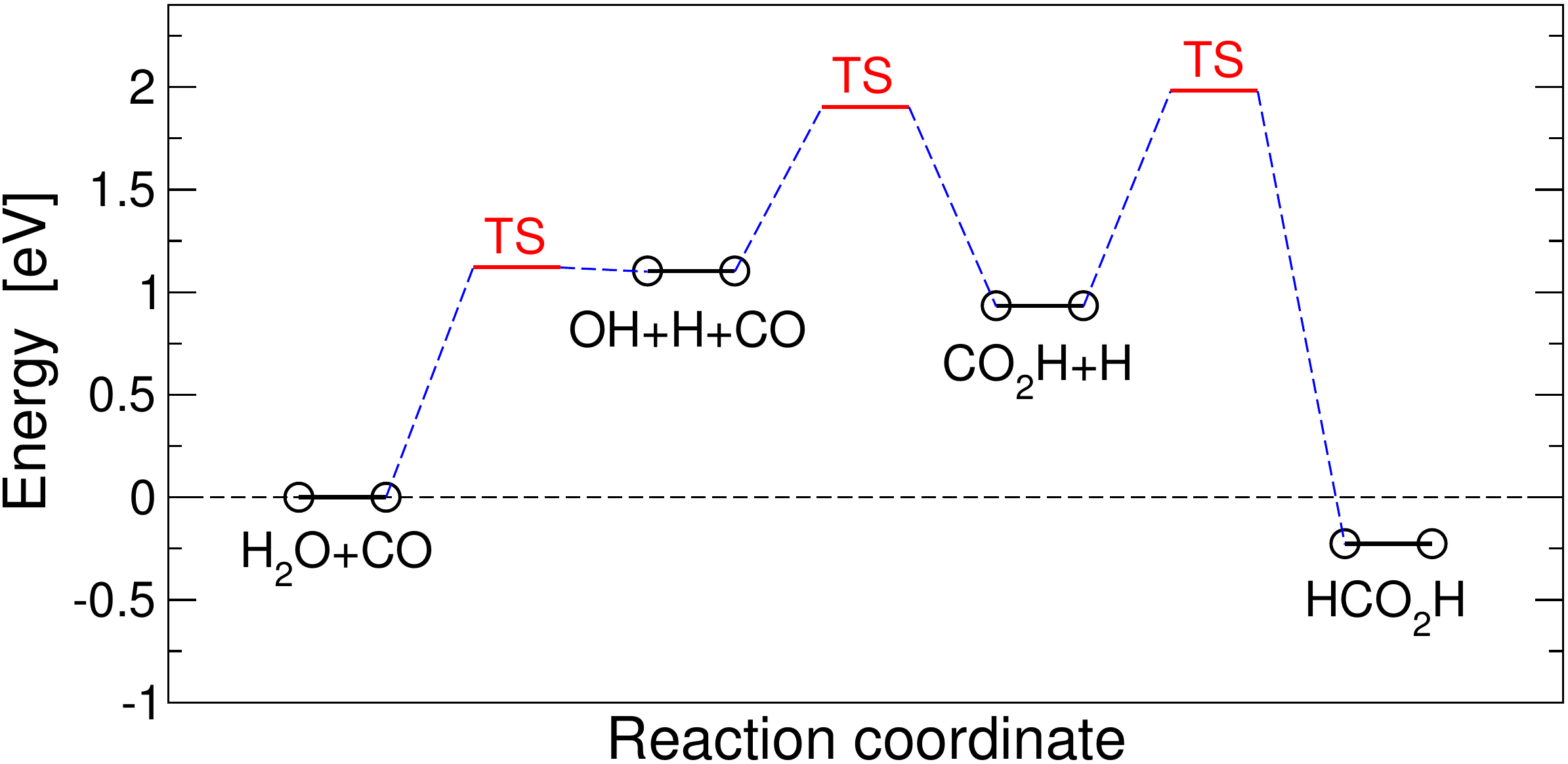}
\caption{\label{fig:energy}Energy of the stable and transition states
(TS) along the {\reactionFA} reaction.}
\end{figure}

We now use our \emph{ab initio} transition-state search to find the
energetically most favorable pathway (i.e.\ lowest energy barriers) for our
proposed reaction pathway.  Results for the structures of reactants, stable
states, and products are depicted in Fig.~\ref{fig:states} and the energy
profile along the entire reaction is plotted in Fig.~\ref{fig:energy}. The
first step of the reaction is the endothermic water dissociation H$_2$O+CO
$\rightarrow$ OH+H+CO. We have previously calculated its reaction barrier
(1.09~eV) and confirmed the separate OH (bound to the open-metal site) and H
(bound to the O of the linker) experimentally inside {\mof} above
150~$^\circ$C.\cite{Tan_2014:water_reaction} Thereafter, the reaction proceeds
exothermic via OH+H+CO $\rightarrow$ CO$_2$H+H $\rightarrow$ HCO$_2$H,
resulting in the formation of formic acid adsorbed on the metal centers of
Zn-{\mof}.  Our calculations show that the energy barrier between the states
OH+H+CO and CO$_2$H+H is 0.8~eV, while the barrier between CO$_2$H+H and
HCO$_2$H is 1.04~eV.  The final state, i.e.\ HCO$_2$H, has an energy 0.23~eV
lower than the energy of the initial state H$_2$O+CO, and the formic acid binds
to the metal centers with an energy of 0.68~eV (65.61~kJ/mol), comparable to
the binding of other molecules to
Zn-{\mof}.\cite{Canepa_2013:high-throughput_screening,
Lee_2015:small-molecule_adsorption} Thus, removal of the formic acid product
from the open-metal sites can proceed through simple activation of the sample,
as confirmed in Fig~\ref{desoprtion_fig}.

The overall barrier for the reaction in Fig.~\ref{fig:energy} is
significant and explains why only a small amount of formic acid is produced.
But, that barrier corresponds to the presence of only one water
molecule.  In related work, we show that the barrier to the first step
of the reaction is lowered by 37\% when the water molecules create
clusters above the linkers.\cite{Zuluaga_2016:controlling_water} It is
conceivable that the presence of several CO and/or H$_2$O molecules can
also lower the energy barrier of the {\reactionHFA} reaction. However,
due to the large number of stable geometries and possible paths for the
{\reactionFA} reaction when more than one molecule is involved, a
comprehensive \emph{ab initio} transition-state search becomes
computational prohibitively expensive.

%%%%%%%%%%%%%%%%%%%%%%%%%%%%%%%%%%%%%%%%%%%%%%%%%%%%%%%%%%%%%%%%%%%%%
\section{Conclusions}
%%%%%%%%%%%%%%%%%%%%%%%%%%%%%%%%%%%%%%%%%%%%%%%%%%%%%%%%%%%%%%%%%%%%%

Our experimental and theoretical work confirms that we can use the OH and H
groups---produced by the {\reaction} reaction---to start a new reaction mechanism
catalyzed inside the confined environment of the Zn-{\mof} channels through water
dissociation  and produce formic acid via {\reactionFA}. This
discovery has several important consequences: It opens the door to a new set of
catalytic reactions inside a well controlled system ({\mof}), it provides a
proof-of-principle that a recovery of the adsorption capacity and structural
stability of Zn-{\mof} is possible after exposure to water, and finally it
produces the important medical/industrial feedstock formic acid.

%%%%%%%%%%%%%%%%%%%%%%%%%%%%%%%%%%%%%%%%%%%%%%%%%%%%%%%%%%%%%%%%%%%%%
\section{Acknowledgements}
%%%%%%%%%%%%%%%%%%%%%%%%%%%%%%%%%%%%%%%%%%%%%%%%%%%%%%%%%%%%%%%%%%%%%

This work was supported by DOE Grant No.\ DE--FG02--08ER46491.
Furthermore, this research used computational resources of the OLCF at
ORNL, which is supported by DOE grant DE--AC05--00OR22725.

%%%%%%%%%%%%%%%%%%%%%%%%%%%%%%%%%%%%%%%%%%%%%%%%%%%%%%%%%%%%%%%%%%%%%%%%
\bibliography{references,biblio}
%\bibliographystyle{rsc}
%%%%%%%%%%%%%%%%%%%%%%%%%%%%%%%%%%%%%%%%%%%%%%%%%%%%%%%%%%%%%%%%%%%%%%%%

%%%%%%%%%%%%%%%%%%%%%%%%%%%%%%%%%%%%%%%%%%%%%%%%%%%%%%%%%%%%%%%%%%%%%%%%
\end{document}